\documentclass[
aps,prl,reprint,amsmath,amssymb,showpacs,nobibnotes
]{revtex4-1}

\usepackage{graphicx}
\usepackage{subfigure}
\usepackage{dcolumn}
\usepackage{bm}

\usepackage[colorlinks]{hyperref}
\usepackage{tabularx}
\usepackage{graphicx,psfrag}
\usepackage{amssymb,bm,tensor}
\usepackage{xcolor}
\usepackage{amsmath}
\usepackage[varg]{txfonts}
\usepackage{enumerate}
\usepackage[normalem]{ulem}
\usepackage{bm}
\usepackage[T1]{fontenc}
\usepackage{orcidlink}
\usepackage{braket}

\usepackage{scalerel,stackengine}
\stackMath
\newcommand\reallywidehat[1]{%
	\savestack{\tmpbox}{\stretchto{%
			\scaleto{%
				\scalerel*[\widthof{\ensuremath{#1}}]{\kern-.6pt\bigwedge\kern-.6pt}%
				{\rule[-\textheight/2]{1ex}{\textheight}}
			}{\textheight}%
		}{0.5ex}}%
	\stackon[1pt]{#1}{\tmpbox}%
}

\newcommand{\rrr}[1]{\vskip 0.2cm \noindent{\bf #1} ---}

\begin{document}
	\title{Relative-Entropy Flow Across the Page Transition}
	\author{Minghui Yu \orcidlink{0000-0003-2257-477X}}
	\affiliation{Department of Physics, College of Sciences, Shanghai University, 99 Shangda Road, 200444 Shanghai, China}

\author{Xian-Hui Ge \orcidlink{0000-0001-6228-0376}}
	\email{gexh@shu.edu.cn}\thanks{Corresponding author}
	\affiliation{Department of Physics, College of Sciences, Shanghai University, 99 Shangda Road, 200444 Shanghai, China}
    
\begin{abstract}
		We develop a computational framework for the quantum relative entropy based on the replica trick and apply it to analyze the evaporation process method of Jackiw-Teitelboim black holes. We show that the relative entropies between the exact radiation state and the Hawking saddle state and the replica wormholes saddle state quantify information-theoretic costs inherent to coarse-grained semiclassical descriptions. During evaporation, the relative entropy for the Hawking saddle increases monotonically, reflecting a progressively higher thermodynamic cost of sustaining the Hawking description. Conversely, the relative entropy associated with the replica wormholes saddle decreases, which indicates that the description of connected geometries becomes progressively more accurate. Moreover, the sign of the difference between the two quantities reverses precisely at the Page time, providing an information-theoretic diagnosis for the transition between competing gravitational saddles. Our results recast the Page transition as a flow of relative entropy costs between competing semi-classical descriptions, revealing an information-theoretic structure underlying the emergence of effective irreversibility in coarse-grained black hole evaporation.

\end{abstract}
	
\maketitle
	
\rrr{Introduction} The black hole (BH) information paradox \cite{IL} lies at the intersection of quantum mechanics (QM), general relativity (GR), quantum information (QI) and thermodynamics. Hawking's semi-classical calculations \cite{HR} demonstrate that BHs evaporation transforms an initially pure state into apparently thermal radiation, raising a fundamental tension with unitarity. Major progress has been achieved through holography \cite{holography}, the island formula \cite{islandrule1,islandrule2}, and replica wormholes \cite{replicawormholes,review}. These developments reproduce the Page curve \cite{pc1,pc2} through a competition between gravitational saddles: the Hawking saddle dominates at early times, while the replica wormholes saddle becomes dominant after the Page time through non-perturbative gravitational contributions, thereby recovering unitary Page curve behavior.

\par Despite this progress, the dynamical and information-theoretic structure underlying the transition between competing gravitational saddles remains poorly understood \cite{zeno}. Existing studies have primarily focused on identifying the dominant saddle responsible for the Page curve \cite{replica1,replica2}, leaving the physical meaning of the transition less explored. In particular, it remains unclear whether an \emph{effective arrow-of-time-like structure} can emerge from competing semiclassical coarse-grained descriptions. Here we show that quantum relative entropy (RE) provides a natural probe of this structure. At the microscopic level,
\begin{equation}
  S(\rho || \sigma) \equiv  \text{Tr} (\rho \log \rho -\rho \log \sigma ),  \label{RE}  
\end{equation}
measures the distinguishability between quantum states. When the exact radiation state is represented by a restricted semiclassical description, part of the microscopic information is discarded, giving rise to an information-theoretic coarse-graining cost. In our framework, the Hawking saddle and replica wormholes saddles provide two such coarse-grained representations of the same radiation state, whose REs therefore quantify the \emph{costs} of the corresponding effective descriptions.

\par This coarse-graining picture connects the saddle transition to QI and non-equilibrium statistical mechanics. The data-processing inequality,
\begin{equation}
    S(\rho|| \sigma) \geq S\Big(\Phi(\rho)|| \Phi(\sigma)\Big),  \label{DPI}
\end{equation}
for a completely positive trace-preserving (CPTP) map $\Phi$, expresses the loss of distinguishability under coarse-graining \cite{cptp}. It therefore provides an information-theoretic basis for interpreting RE as a coarse-graining cost, although the time dependence of the REs follows from the explicit gravitational saddle dynamics rather than from the inequality itself. A complementary thermodynamic perspective is provided by fluctuation relations: the Crooks theorem \cite{crooks} relates the asymmetry between forward and time-reversed trajectories to stochastic entropy production,
\begin{equation}
\frac{P_F(x)}{P_R(x)} = e^{\Sigma(x)},  \label{crooks formula}  
\end{equation}
where $\Sigma[x]$ denotes the stochastic entropy production \cite{crooks}. Here, $\Sigma[x]$ and the RE characterize irreversibility at different levels—the former for trajectories and the latter for quantum states and their coarse-grained representations. A detailed discussion of this distinction and the conditions for a direct gravitational analogue is given in appendix A.

\par To investigate this connection in a gravitational system, we employ the PSSY model \cite{replica1} as the theoretical framework for this study \footnote{For conciseness, the PSSY model as well as the derivation of its R\'enyi entropy and Page curves is detailed in \cite{replica1}.}. This model is a two-dimensional toy model that composed of an End-of-the-World (EoW) brane with $k$ flavors in Jackiw–Teitelboim (JT) gravity \cite{jtgravity1,jtgravity2}. It captures essential features of BH evaporation, including quantum entanglement, information recovery, and replica saddle competition, while allowing explicit calculations of the gravitational path integral (GPI) through the replica trick \cite{replica3}. Based on this framework, we develop a replica trick for calculating the Petz–R\'enyi relative entropy (PRRE) between the exact radiation state and semi-classical saddles.
\begin{equation}
    S_n(\rho || \sigma) = \frac{1}{n-1} \log \text{Tr}[(\rho^n \sigma^{1-n})],  \label{PRRE}
\end{equation}
with the RE \eqref{RE} recovered in the limit $n \to 1$. We interpret the Hawking saddle and replica wormholes saddle as two distinct coarse-grained representations of the radiation state and quantify the associated information costs. We find a monotonic RE flow across the Page transition: the description of Hawking saddle becomes increasingly costly, while the description of replica wormholes saddle becomes progressively favored. These findings establish a connection between gravitational saddle dynamics, QI theory, and the emergence of irreversible behavior in semi-classical coarse-grained description.

\rrr{RE of saddles from replica geometry} To quantify the validity of different semiclassical descriptions during BH evaporation, we introduce the RE between the exact radiation state and reference states of saddles. The RE provides a natural measure of the distinguishability between two quantum states. More importantly, in the present gravitational context, it quantifies the information-theoretic cost required to approximate an exact quantum state by an effective semiclassical description.

\par In the PSSY model, the exact density matrix of radiation  is obtained by tracing out the degree of freedom of BH from the whole system  $\ket{\Psi}$ that generated by the GPI, namely, the reduced radiation state can be expressed as $\rho_{\text{ext}}=\text{Tr}_{\text{BH}}\ket{\Psi}\bra{\Psi}$.
This density matrix represents the unitary radiation state containing all correlations generated during the evaporation process. In contrast, the semiclassical GPI admits two dominant saddle configurations: the Hawking saddle and the replica wormholes saddle. We therefore introduce two saddle-resolved reference states, $\sigma_{\text{Haw}}$ and $\sigma_{\text{WH}}$, obtained by projecting the GPI onto the Hawking saddle and replica wormholes saddle sectors, respectively. 
\begin{equation}
    \sigma_{\text{Haw}}= \rho_{\text{ext}}^{\text{Haw}}, \quad 
    \sigma_{\text{WH}}= \rho_{\text{ext}}^{\text{WH}}. \label{two saddles}
\end{equation}
It is important to emphasize that these reference states are not independent microscopic states. Instead, they represents two different \emph{coarse-grained semiclassical descriptions} of the same radiation system. Therefore, the RE between the state $\rho_{\text{ext}}$ and these reference states does not simply measure the abstract distinguishability between two states, but quantifies the \emph{information cost} associated with representing the exact radiation state through a given semiclassical approximation.  Moreover, it is essential to note that these states are normalized under the appropriate boundary conditions, and we assume 
 \begin{equation}
    \text{supp}  (\rho_{\text{ext}})\subseteq \text{supp}(\sigma), \label{support}
 \end{equation}
to ensure convergence of all subsequent calculations.

\par For completeness, we first recall that the conventional replica construction for the PSSY model involves $n$ identical copies of the radiation density matrix \cite{replica1}. By appropriately gluing these replicas according to the boundary conditions, the EE can be obtained by tracing this replica: $S_{\text{vN}} = \lim\limits_{n\to1} \frac{1}{n-1} \log [\text{Tr}(\rho_{\text{ext}}^n)]$. Finally, the result reproduce an unitary Page curve \footnote{The calculation of R\'enyi entropy for the PSSY model is shown in \cite{replica3}.}
\begin{equation}
\begin{split}
S_{\text{vN}}            &\simeq \left \{
              \begin{array}{lr} \log k , \qquad \qquad \ \  \qquad t<t_{\text{Page}}. &\\
              \lim\limits_{n \to 1}S_{\text{BH}}^{(n)} =S_{\text{BH}}, \qquad \ t>t_{\text{Page}}. &
              \end{array} \right.  \label{vn entropy}
\end{split}
\end{equation}
where  $S_{\text{BH}}^{(n)} = \frac{1}{1-n} \log \Big[\frac{Z_n}{Z_1^n} \Big]$ is the R\'enyi BH entropy. However, the current calculation of RE requires a generalized replica construction because it involves \emph{two distinct} density matrices. We now concentrate on the PRRE \eqref{PRRE}, whose $n \to 1$limit gives the RE \eqref{RE}. In the present work, we choose the exact radiation state as the first argument $\rho=\rho_{\text{ext}}$, while the reference state is selected as either the Hawking saddle $\sigma_{\text{Haw}}$ or the replica wormholes saddle $\sigma_{\text{WH}}$. These two choices correspond to two competing semiclassical descriptions of the same quantum radiation state.  Secondly, rather than simply quantifying state distinguishability, based on the definition \eqref{two saddles}, the RE in this context measures the \emph{information-theoretic cost} of approximating the exact radiation state by a semiclassical effective description. Specifically, the RE $S(\rho_{\text{ext}}||\sigma_{\text{Haw}})$ characterizes the cost of maintaining the Hawking semiclassical description when the exact radiation state $\rho_{\text{ext}}$ is considered as the underlying quantum reality. Since the Hawking saddle neglects the non-perturbative correlations responsible for information recovery, this quantity measures the accumulated information cost induced by the Hawking coarse-graining. Consequently, an increasing/decreasing value of this RE therefore indicates that the Hawking description becomes increasingly/decreasingly inefficient as an effective representation of the exact state. In contrast,  the other RE $S(\rho_{\text{ext}}||\sigma_{\text{WH}})$ measures the information cost associated with describing the exact radiation state through the replica wormholes saddle. Since the replica wormholes saddle incorporates the non-perturbative correlations required for recovering unitarity, this quantity evaluates how accurately the replica wormhole geometry captures the underlying quantum state. The comparison between these two REs determines which semiclassical description provides a more precise coarse-grained representation of the radiation state. Moreover, to quantify this competition, we define the difference of these two REs
\begin{equation}
  \Delta S \equiv S(\rho_{\text{ext}} || \sigma_{\text{Haw}}) - S(\rho_{\text{ext}} || \sigma_{\text{WH}}) \label{difference RE}
\end{equation}
serves as an information-theoretic diagnostic for the transition between competing gravitational saddles. The sign of $\Delta S$ determines which saddle incurs a smaller coarse-graining cost, thereby providing a quantitative characterization of the transition between competing semiclassical descriptions.

\par Using the generalized replica construction, the detailed derivation of the two REs is presented in the appendix B. Here we summarize the resulting analytical expressions. The expression of Hawking saddle RE is
\begin{equation}
      S(\rho_{\text{ext}} || \sigma_{\text{Haw}})= -S_{\text{vN}}(\rho_{\text{ext}}) + \log k. \label{RE Haw}
\end{equation}
This expression separates into two contributions: the EE of the exact radiation state and the coarse-graining contribution associated with the inaccessible degrees of freedom encoded by the EoW brane index $k$. While for the replica wormholes saddles, the corresponding expression has the following form
\begin{equation}
     S(\rho_{\text{ext}} || \sigma_{\text{WH}})= -S_{\text{vN}}(\rho_{\text{ext}}) +S_{\text{BH}}.  \label{RE WH}
\end{equation}
Here the coarse-graining contribution is replaced by the BH entropy, reflecting the additional correlations incorporated by the replica wormholes geometry. Consequently, the difference   between these two REs \eqref{difference RE} is independent of the EE of the exact radiation state and is determined solely by the competition between the two semiclassical saddle contributions
\begin{equation}
    \Delta S \equiv S(\rho_{\text{ext}} || \sigma_{\text{Haw}}) - S(\rho_{\text{ext}} || \sigma_{\text{WH}}) = \log k - S_{\text{BH}}.  \label{order parameter}
\end{equation}
In the following contents, we analyze the evolution of quantities and reveal the emergence of effective irreversible behavior in semiclassical coarse-grained descriptions of BH evaporation.

\begin{figure}[t]
\centering
\includegraphics[scale=0.70]{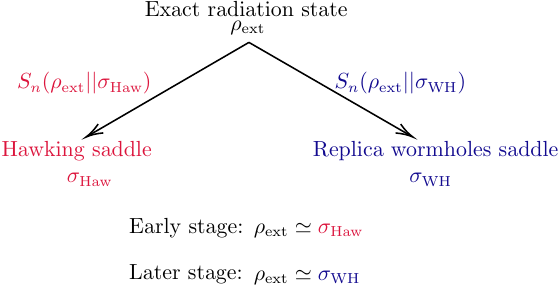}
\caption{Schematic illustration of the RE between the exact radiation state and saddle-resolved semiclassical descriptions.}
\label{states fig}
\end{figure}

\rrr{RE flow and coarse-graining cost across the Page transition}  We now present the time-dependent evolution of RE \eqref{RE Haw} \eqref{RE WH} and physical implications for the transition between competing semiclassical gravitational saddles. The central purpose of this content is that the two reference states $\sigma_{\text{Haw}}$ and $\sigma_{\text{WH}}$, should not be interpreted as independent microscopic states. Instead, they represent two distinct coarse-grained descriptions of the same exact radiation state $\rho_{\text{ext}}$. Therefore, the RE provides a new interpretation for characterizing the information-theoretic cost incurred by distinct semi-classical saddle configurations in the description of approximately exact quantum states. To illustrate this interpretation, FIG. \ref{states fig} presents a schematic representation of the three quantum states involved in our analysis. The exact radiation state $\rho_{\text{ext}}$ is obtained from the unitary evolution of the total system, while $\sigma_{\text{Haw}}$ and $\sigma_{\text{WH}}$ correspond to the radiation states reconstructed from the Hawking saddle and replica wormholes saddle, respectively. The REs \eqref{RE Haw} \eqref{RE WH} therefore quantify the coarse-graining costs of these two effective descriptions.

\par
\begin{figure*}[t]
\centering
\subfigure[\scriptsize{}]{\label{relative1}
\includegraphics[scale=0.3]{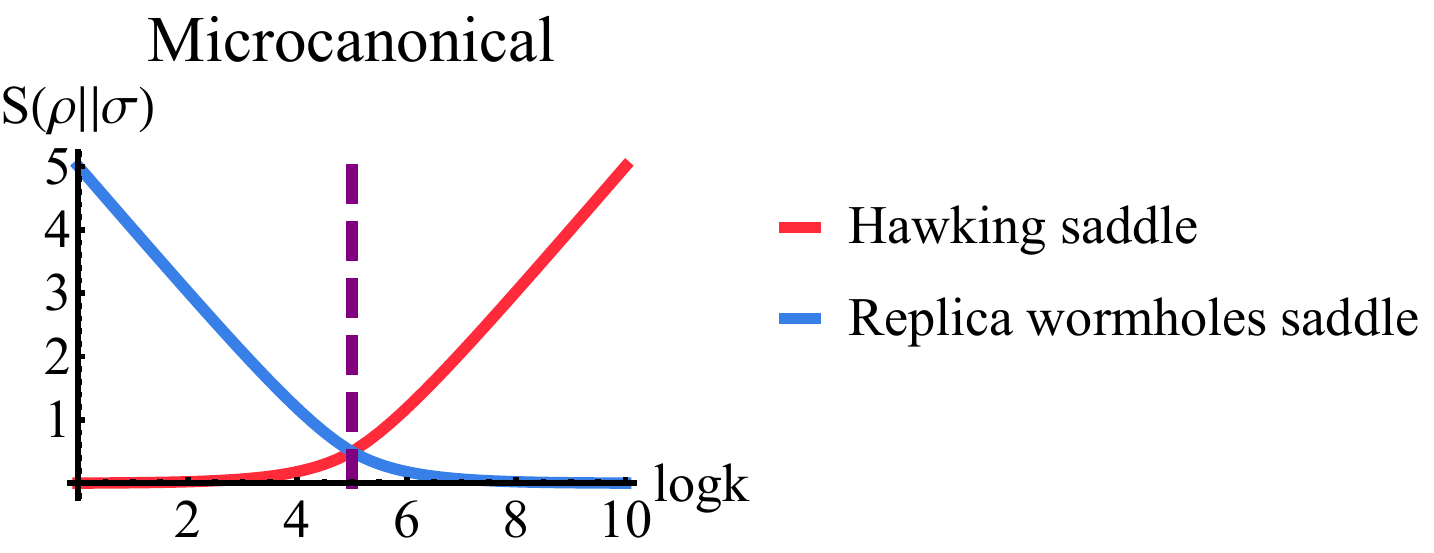}
}
\qquad \qquad \quad
\subfigure[\scriptsize{}]{\label{relative2}
\includegraphics[scale=0.41]{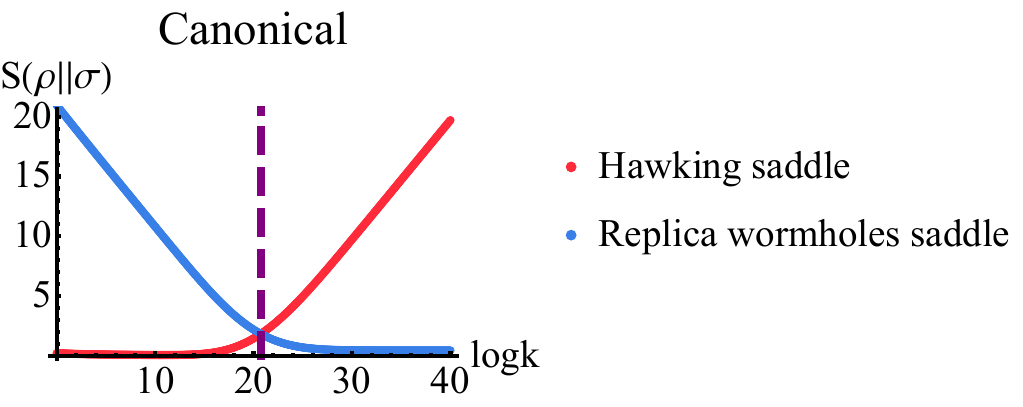}
}
\caption{The time-dependent evolution of RE. The left panel displays the results for micro-canonical ensemble, while the right panel shows the results of the canonical ensemble.  Red curves are the  RE of the Hawking saddle \eqref{RE Haw ME} \eqref{RE Haw CE}. Blue curves are the RE of the replica wormholes saddle \eqref{RE WH ME} \eqref{RE WH CE}. 
Purple denotes the Page time $\Big($$t_{\text{Page}}^{\text{ME}}=S_0=5$ for the micro-canonical ensemble. $t_{\text{Page}}^{\text{CE}}=20.739$  for the canonical ensemble,  which defined by $\Delta S(t_{\text{Page}})=0$ $\Big)$. Notably, the RE of Hawking saddle starts from a \emph{non-zero} value and keeps a continuous monotonic increase. See appendix C and D for more detailed explanations. }
\label{RE figures}
\end{figure*}

 The PSSY model is characterized by two different boundary conditions: fixed energy $E$ and fixed (inverse) temperature $\beta$, which corresponds to the micro-canonical and canonical ensemble, respectively. In both cases, REs between the exact radiation state and the saddle-resolved reference states can be evaluated analytically or numerically. For the micro-canonical ensemble, REs \eqref{RE Haw} \eqref{RE WH} is given by \footnote{See appendix C for the detailed derivation.}
\begin{equation}
\begin{split}
S_{\text{ME}}(\rho_{\text{ext}}||\sigma_{\text{Haw}})            &= \left \{
              \begin{array}{lr} \log \frac{k}{t} -S_0 + \frac{t}{2} , \quad  0\leq t\leq1. &\\
              \log k -S_0 + \frac{1}{2t}, \ \  \qquad t\geq1. &
              \end{array} \right.  \label{RE Haw ME}
\end{split}
\end{equation}
and
\begin{equation}
\begin{split}
S_{\text{ME}}(\rho_{\text{ext}}||\sigma_{\text{WH}})            &= \left \{
              \begin{array}{lr} -\log t + \frac{t}{2} , \quad  0\leq t\leq1. &\\
              \frac{1}{2t}, \qquad  \qquad \quad \ \ \ \  t\geq1. &
              \end{array} \right.  \label{RE WH ME}
\end{split}
\end{equation}
Here $t$ is a time-scale factor and is set by $t\equiv \frac{k}{e^{S_0}}$. The Page time is defined by $t_{\text{Page}}=1$. The behavior of RE of Hawking saddle is depicted in FIG. \ref{relative1}. While for the canonical ensemble, the RE is expressed as follows
\begin{equation}
S_{\text{CE}}(\rho_{\text{ext}}||\sigma_{\text{Haw}}) = \int_0^{s_k} ds \cdot \tilde{\rho}(s) \big( \lambda_0 +w(s)\big) \log\big( \lambda_0+w(s)\big)+\log k. \label{RE Haw CE}
\end{equation}
\begin{equation}
S_{\text{CE}}(\rho_{\text{ext}}||\sigma_{\text{WH}}) = \int_0^{s_k} ds \cdot \tilde{\rho}(s) \big( \lambda_0 +w(s)\big) \log\big( \lambda_0+w(s)\big)+ S_{\text{BH}}. \label{RE WH CE}
\end{equation}
These two integrals admit straightforward numerical calculation. See appendix D for the concrete calculation under the canonical ensemble. Consequently, we plot the evolution curve of RE of replica wormholes saddle as a function of $\log k(=\log t+S_0)$ with the parameter $S_0=1$, $\beta=2\pi$, and $\mu=5$ in FIG. \ref{relative2}. 

\par These results in FIG. \ref{RE figures} illustrates the highly structured behavior of RE through the BH evaporation process. The RE with respect to the Hawking saddle, $S(\rho_{\text{ext}}||\sigma_{\text{Haw}})$, almost to vanishes at early times and increases monotonically, indicating an increasing information cost for maintaining the Hawking semiclassical description. In contrast, the RE for the replica wormholes saddle, $S(\rho_{\text{ext}}||\sigma_{\text{WH}})$, initiates at a large value of ${\cal O}(S_{\text{BH}})$ and asymptotically decays to zero at late times. This behavior shows
that the description of connected geometries becomes progressively closer to the exact radiation state. Importantly, this evolution should not be interpreted as a loss of fundamental QI. The exact state $\rho_{\text{ext}}$ always follows unitary evolution. Instead, the RE flow defined by $S(\rho_{\text{ext}}||\sigma_{\text{Haw}})$ and $S(\rho_{\text{ext}}||\sigma_{\text{WH}})$ measures how the efficiency of different semiclassical coarse-grained descriptions changes during evaporation. To quantitatively characterize the transition between these two effective descriptions, we use the difference RE defined by \eqref{order parameter}. The evolution of this quantity is plotted in FIG. \ref{order parameter}. It evolves monotonically and \emph{vanishes} exactly at the Page time. The sign of $\Delta S$ determines the dominant semi-classical saddle configuration approximating the exact radiation state: a negative value indicates that the Hawking saddle provides a more accurate description, whereas a positive value signifies dominance of the replica wormholes saddle. Therefore, $\Delta S$ functions naturally as an information-theoretic \emph{diagnosis}  characterizing the transition between competing semiclassical saddles. Unlike the conventional Page curve, which describes the behavior of EE, this quantity directly quantifies the relative accuracy of different semiclassical descriptions. 

\begin{figure}[t]
\centering
\subfigure[\scriptsize{}]{\label{parameter1}
\includegraphics[scale=0.25]{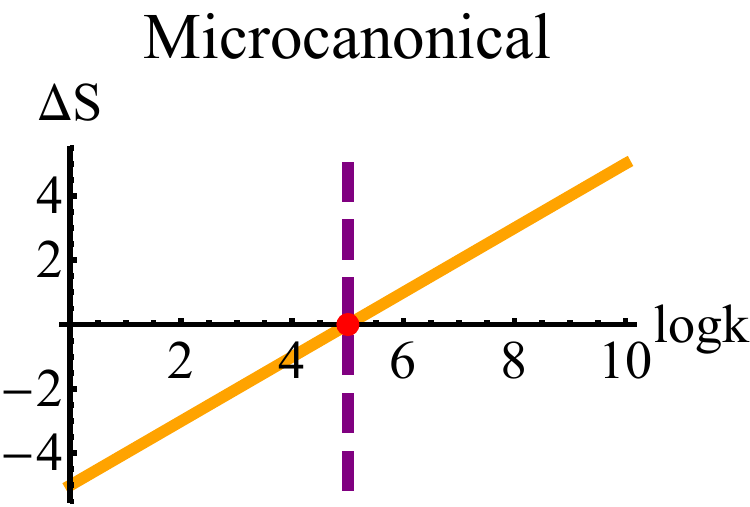}
}
\qquad
\subfigure[\scriptsize{}]{\label{parameter2}
\includegraphics[scale=0.25]{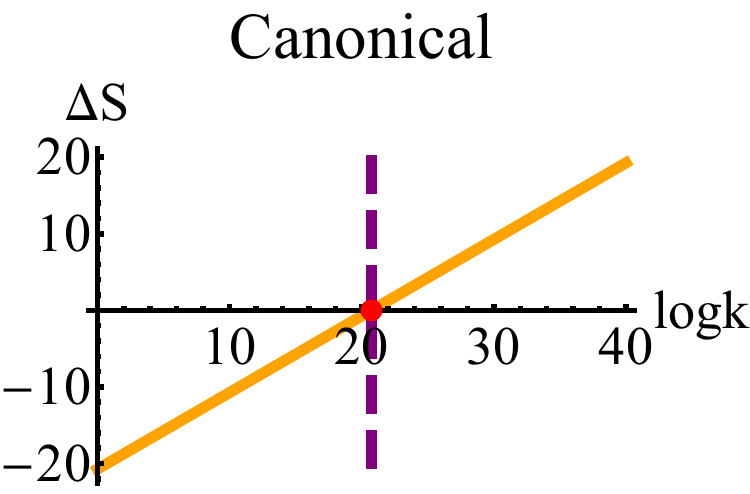}
}
\caption{The difference in RE defined by \eqref{order parameter} under two ensembles. The sign change of $\Delta S$ provides an information-theoretic diagnostic of the saddle transition.}
\label{order parameter}
\end{figure}

\par The physical origin of the RE flow can be understood from the perspective of coarse-graining. Importantly, the semiclassical nature of the Hawking saddle does not by itself imply irreversibility or an arrow of time. Rather, the effective irreversible structure arises from the specific coarse-graining involved in replacing the exact radiation state by a semiclassical description. The Hawking saddle does not represent an independent microscopic quantum state; instead, it provides an effective description that neglects certain non-perturbative correlations encoded in the exact radiation state. We may therefore formally represent this coarse-graining by an effective CPTP map $\Phi$ satisfying
\begin{equation}
    \Phi(\rho_{\text{ext}}) \mapsto \sigma_{\text{Haw}}. \label{map}
\end{equation} 
Based to the quantum data processing inequality \eqref{DPI} and using \eqref{map}, we have
\begin{equation}
    S(\rho_{\text{ext}}|| \sigma_{\text{Haw}}) =   S\bigg(\rho_{\text{ext}}|| \Phi(\rho_{\text{ext}})\bigg). \label{coares graining}
\end{equation}
In this sense, this RE quantifies the information-theoretic cost incurred when approximating the exact quantum radiation state with the Hawking semi-classical description. A higher value signifies that a greater amount of microscopic information is necessary to reconstruct the underlying unitary evolution. In particular, the data processing inequality provides a general information-theoretic foundation for this interpretation. Then the effective description obtained after coarse-graining can not retain more information about microscopic differences than the original QI. Nevertheless, the data processing inequality alone does not govern the time evolution of the RE under consideration. Rather, the monotonic increase of $S(\rho_{\text{ext}}||\sigma_{\text{Haw}})$ arises explicitly from the entropy dynamics prescribed by the PSSY model and the Page-curve evolution established earlier. Consequently, the primary function of the data processing inequality is to substantiate the interpretation of the RE as a \emph{quantitative measure of coarse-graining cost}, not to serve as the dynamical origin of its monotonic behavior. The connection to non-equilibrium thermodynamics can be elucidated via an analogy with the fluctuation theorems \eqref{crooks formula}. Specifically, within the framework of the Crooks fluctuation theorem, entropy production serves as a quantitative measure of the statistical asymmetry between forward and reverse dynamical trajectories, thereby characterizing the thermodynamic cost associated with irreversible processes. In the present gravitational context, however, $S(\rho_{\text{ext}}||\sigma_{\text{Haw}})$ does not represent trajectory-level entropy production and is not directly identifiable with the entropy production term appearing in the Crooks relation. Rather, $S(\rho_{\text{ext}}||\sigma_{\text{Haw}})$ assumes a formally analogous role as an information-theoretic quantifier of the descriptive cost incurred when adopting an effective coarse-grained description, as opposed to the exact unitary evolution of the underlying quantum system. Consequently, the increasing Hawking saddle RE therefore indicates that the semiclassical Hawking description becomes progressively less efficient as an approximation of the underlying quantum state. This behavior reveals an emergent \emph{effective irreversible} structure in the evolution of coarse-grained gravitational descriptions—despite the preservation of unitarity at the fundamental microscopic level.

\par We emphasize that the relative entropy introduced herein is conceptually \emph{analogous} to the entropy production measure featured in Crooks fluctuation relations, as both quantify asymmetry arising from an effective dynamical description. Nevertheless, the two quantities are distinct: the present relative entropy quantifies the discrepancy between an exact quantum state and a coarse-grained semiclassical saddle-point approximation, whereas Crooks entropy production is rigorously defined in terms of the ratio of \emph{forward} and \emph{reverse} trajectory probabilities (see appendix A).

\rrr{Discussion and Conclusion} In summary,  we develop a replica-based framework for quantifying quantum RE in gravitational systems and apply it to the PSSY model of BH evaporation. We show that the Page transition can be reformulated as a RE flow between two competing coarse-grained semiclassical descriptions. The Hawking saddle and replica wormholes saddle corresponds to two different approximations of the exact radiation state $\rho_{\text{ext}}$, characterized by distinct information-theoretic costs. During evaporation, the RE  $S(\rho_{\text{ext}}||\sigma_{\text{Haw}})$ associated with the Hawking description increases, while that  $S(\rho_{\text{ext}}||\sigma_{\text{WH}})$  associated with the replica wormholes description decreases, and their difference changes sign at the Page time, providing an information-theoretic criterion for the saddle transition. Importantly, the observed RE flow does not indicate a breakdown of \emph{microscopic unitarity}. Instead, it reflects the emergence of \emph{an effective irreversible structure} when the exact quantum evolution is represented through coarse-grained semiclassical descriptions. The increasing RE with respect to the Hawking saddle quantifies the growing cost of neglecting correlations required for information recovery, whereas the replica wormholes saddle provides an increasingly efficient description of these correlations at late times. In this sense, the RE flow plays a role \emph{analogous} to entropy production in non-equilibrium thermodynamics, although a direct identification would require an explicit microscopic forward–reverse gravitational protocol. Our results establish a connection between gravitational saddle dynamics and QI theory by showing that the Page transition is not only a change of dominant replica geometry but also a transition between competing information costs. Extending this framework to more realistic systems, such as strongly coupled Sachdev-Ye-Kitaev model system \cite{syk1,syk2,syk3} or quantum circuits, thereby enabling further investigation of RE flows and emergent irreversible structures in holographic dynamics.

\rrr{Acknowledgments} We would like to thank Zhenbin Yang Rene Meyer, Johanna Erdmenger, SangJin Sin, and Zhuoyu Xian for helpful suggestions.The study was partially supported by NSFC, China (Grant No.12275166, No.12675074 and No.12547142).

\begin{widetext}
\renewcommand{\theequation}{S\arabic{equation}}
\setcounter{equation}{0}
\renewcommand{\thefigure}{S\arabic{figure}}
\setcounter{figure}{0}

\newpage

\section{Supplemental Material}
\renewcommand{\theequation}{S\arabic{equation}}
\setcounter{equation}{0}
\renewcommand{\thefigure}{S\arabic{figure}}
\setcounter{figure}{0}

\subsection{A.  Relative entropy and irreversibility}
\par In this appendix, we establish the connection between the RE flow found in this work and the emergence of an effective irreversible structure. The key point is that the REs considered here quantify the information-theoretic costs of representing the exact radiation state by competing coarse-grained semiclassical descriptions. Their time-dependent flow therefore provides a measure of the directional change in the information required to describe the same underlying quantum state within these effective descriptions..

\par The two REs considered in this work, $S(\rho_{\text{ext}}|| \sigma_{\text{Haw}})$ and $S(\rho_{\text{ext}}|| \sigma_{\text{WH}})$ measure the mismatch between the exact radiation state $\rho_{\text{ext}}$ and the Hawking and replica wormholes descriptions, respectively. A smaller RE means that the corresponding saddle provides a more economical description of the exact state, whereas a larger RE indicates a larger information-theoretic cost associated with the corresponding coarse-graining. During evaporation, we find $S(\rho_{\text{ext}}|| \sigma_{\text{Haw}})$ increases while $S(\rho_{\text{ext}}|| \sigma_{\text{WH}})$ decreases. Therefore, the two effective descriptions acquire opposite directional flows of information-theoretic cost. The Hawking description progressively loses its ability to account for the correlations encoded in the exact radiation state, while the replica wormholes description progressively incorporates these correlations. Their RE costs cross at the Page time. The Page transition can therefore be interpreted as a directional reorganization of information between competing effective descriptions, rather than merely as a change in the dominant gravitational saddle.

\par This provides the information-theoretic origin of the effective irreversibility discussed in the main text. Importantly, the irreversibility does not arise from semi-classicality itself. Rather, it arises because the exact microscopic state is represented through restricted semiclassical descriptions that discard different portions of the underlying information. Once such a coarse-grained description is applied, the corresponding information cost acquires a definite ``time of arrow'': the cost of the Hawking description increases, whereas that of the replica wormholes description decreases. This directional flow defines an effective irreversible structure at the level of the coarse-grained description, even though the underlying microscopic quantum evolution remains unitary. In this sense, the Page transition marks not only the exchange of gravitational saddle dominance but also the exchange of informational efficiency between two effective descriptions of the same quantum state. The data processing inequality  gives a general information-theoretic foundation for this interpretation. For a CPTP map $\Phi$, we have the relation \eqref{DPI}. It expresses the fact that coarse-graining cannot increase the distinguishability of quantum states. Accordingly, when a microscopic state is replaced by a effective description, information about the microscopic distinction between states is necessarily lost. In the present gravitational setting, the saddle-dependent REs quantify the resulting information-theoretic mismatch. The data processing inequality itself does not determine their time dependence; the monotonic flows arise from the explicit gravitational dynamics in the PSSY model. It nevertheless provides the general information-theoretic principle that makes the interpretation of these flows as coarse-graining costs natural.

\par The connection to non-equilibrium thermodynamics can then be made through the Crooks fluctuation theorem \cite{crooks}. Based on the theorem \eqref{crooks formula}, Crooks relation identifies irreversibility with a directional asymmetry between a process and its reverse. The RE framework developed here realizes an analogous structure at the \emph{level of effective quantum descriptions}: instead of comparing forward and reverse trajectories, we compare the exact microscopic state with competing coarse-grained semiclassical representations. The corresponding RE flow measures how the information-theoretic cost of these representations changes along the evaporation process. Thus,
\begin{equation}
\text{trajectory asymmetry}
\longleftrightarrow \Sigma[x],
\qquad
\text{coarse-grained state mismatch}
\longleftrightarrow
S(\rho_{\text{ext}}||\sigma),
\end{equation}
with both providing directional measures of irreversibility at different levels of description. Therefore, the RE $S(\rho_{\text{ext}}|| \sigma_{\text{Haw}})$  acts as an information-theoretic cost for the reorganization of the effective description. The RE flow and its crossing provide, in this sense, a state-level analogue of the directional information loss that underlies entropy-production descriptions of non-equilibrium processes. A literal Crooks relation for black-hole evaporation would require an explicit microscopic forward--reverse gravitational protocol and trajectory-level probability distributions, which is beyond the present construction. The present result instead establishes the corresponding information-theoretic structure directly: competing semiclassical coarse-grainings acquire oppositely directed RE flows, and their exchange at the Page time gives rise to an emergent effective irreversibility while remaining fully compatible with microscopic unitarity.

\subsection{B. The explicit derivation for REs}
\par In this appendix, we provide a rigorous and step-by-step derivation of expressions. We star from the PRRE for the Hawking saddle: $S_n(\rho_{\text{ext}}||\sigma_{\text{Haw}}) = \lim\limits_{m\to 1-n} \log [\text{Tr}\big(\rho_{\text{ext}}^n \sigma_{\text{Haw}}^m\big)]$. The core structural component is the trace $\text{Tr}\big(\rho_{\text{ext}}^n \sigma_{\text{Haw}}^m\big)$. It represents the summation of $n$ copies of state $\rho_{\text{ext}}$ and $m$ copies of state $\sigma_{\text{Haw}}$. A schematic illustration of this process is provided in FIG. \ref{trace} below.
\begin{figure}[htb]
\centering
\includegraphics[scale=0.20]{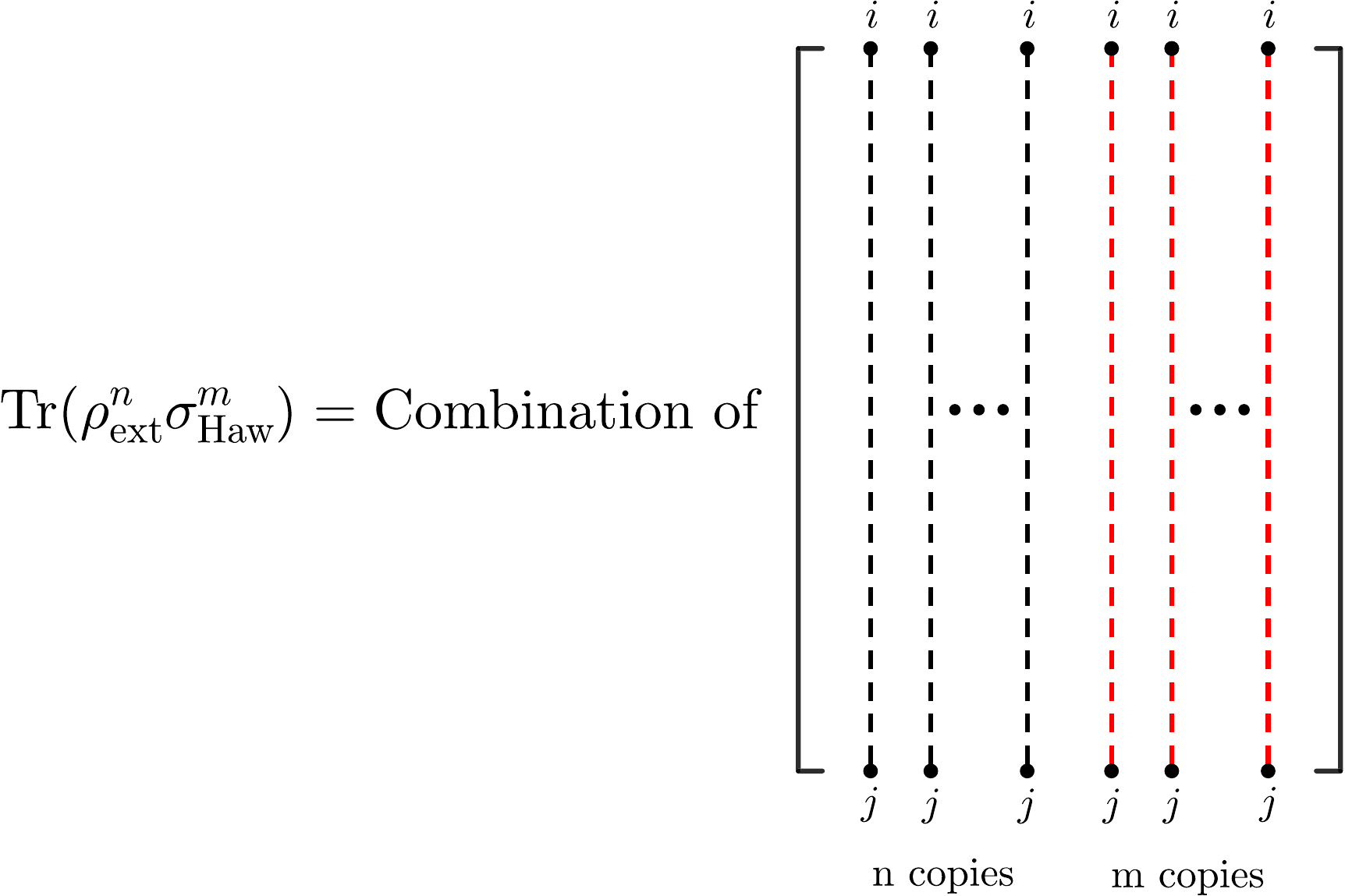}
\caption{A graphical depiction of $\text{Tr}\big(\rho_{\text{ext}}^n \sigma_{\text{Haw}}^m\big)$. To clearly differentiate between the two distinct states, we adopt black to denote state $\rho_{\text{ext}}$ and red to denote $\sigma_{\text{Haw}}$. Each dashed line in the figure denotes the gravitational amplitude $\braket{\psi_i|\psi_j}$. }
\label{trace}
\end{figure}

 It is noted that, due to the fact that distinct boundary conditions associated with the two types of states, the combined result can not yield a configuration in which two states are connected. Namely, red lines and blue lines can not connect independently and can not glue together. It is also noteworthy that the state $\sigma_{\text{Haw}}$ corresponds to the Hawking saddle, which yields a \emph{full disconnected geometry}. Therefore, the trace $\text{Tr}\big(\rho_{\text{ext}}^n \sigma_{\text{Haw}}^m\big)$ can be simplified to the form illustrated in FIG. \ref{replicageometry1}, as supported by the two preceding points.
\begin{figure}[htb]
\centering
\includegraphics[scale=0.3]{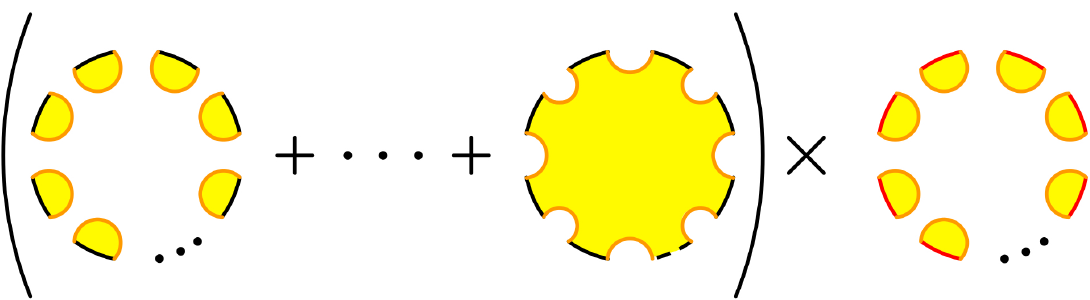}
\caption{Graphical representation of the reduced form of the trace $\text{Tr}\big(\rho_{\text{ext}}^n \sigma_{\text{Haw}}^m\big)$.  }
\label{replicageometry1}
\end{figure}

The first term is nothing but $\text{Tr}(\rho_{\text{ext}}^n)$, whereas the second term corresponds to the sum of Hawking saddle. The latter only result in the \emph{full} disconnected geometry. As previously noted, lines of differing colors are incompatible and can not interconnected. Consequently, the resulting composite configuration becomes a geometric multiplication of the two types. Moreover, there is one last attention. The product resulting from the first term in FIG. \ref{replicageometry1} yields two \emph{full} disconnected geometries, as shown in below.
\begin{figure}[htb]
\centering
\includegraphics[scale=0.3]{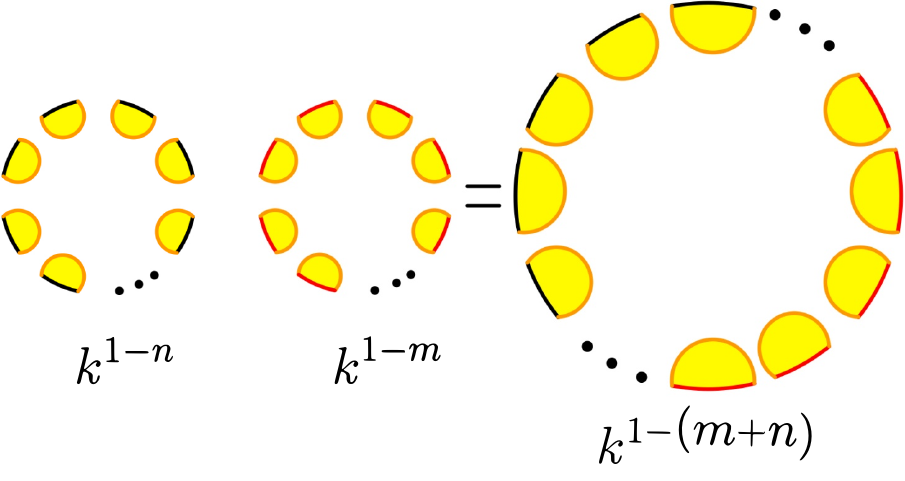}
\caption{The combination of two full disconnected geometries introduces an additional multiplicative factor of $k^{-1}$, as such geometries do not distinguish between the $\rho_{\text{ext}}$ and $\sigma_{\text{Haw}}$ state.}
\label{disconnect}
\end{figure}

Since the union of the two disconnected geometries structures remains topologically disconnected, its contribution is $k^{1-(m+n)} \neq k^{1-n} \cdot k^{1-m}$, rather than the product of their individual contribution. Therefore, a compensation term $k^{-1}$ must be introduced to rectify the result. In light of the foregoing considerations, the expression of the trace  $\text{Tr}\big(\rho_{\text{ext}}^n \sigma_{\text{Haw}}^m\big)$ is given by
\begin{equation}
    \text{Tr}\big(\rho_{\text{ext}}^n \sigma_{\text{Haw}}^m\big)  =   \text{Tr}\big(\rho_{\text{ext}}^n\big) \cdot \text{Tr}\big(\sigma_{\text{Haw}}^m\big) \cdot k^{-1}. \label{traceoverlap}
\end{equation}
For the PSSY model, in the limit of large $k$, the planar topology approximation becomes valid, yielding \cite{replica3}
\begin{equation}
    \text{Tr}\big( \rho_{\text{ext}}^n\big) = \frac{1}{k^{n-1}}  \Bigg[ 1+ C_n^2 \frac{kZ_2}{Z_1^2}+\cdots+\frac{k^{n-1}Z_n}{Z_1^n}  \Bigg]. \label{traceradiation}
\end{equation}
Here $C_n^m$ denoted as the combinatorial number, $Z_n$ is the (replica) partition function. The first and last terms in the expression correspond, respectively, to the contributions from the Hawking saddle and the replica wormholes saddle, i.e., 
\begin{equation}
    \text{Tr}\big(\sigma_{\text{Haw}}^n \big) = \frac{1}{k^{n-1}}, \qquad \text{Tr}\big(\sigma_{\text{WH}}^n \big) = \frac{Z_n}{Z_1^n}.  \label{tracesaddles}
\end{equation}
Therefore, the final expression for PRRE \eqref{PRRE} and RE \eqref{RE} as follows, respectively
\begin{equation}
    \begin{split}
         S_n(\rho_{\text{ext}} || \sigma_{\text{Haw}}) &= \frac{1}{1-n}\log \bigg[\text{Tr}\big(\rho_{\text{ext}}^n\big) \cdot \text{Tr}\big(\sigma_{\text{Haw}}^m\big) \cdot k^{-1} \bigg] \\
         &=-S_n(\rho_{\text{ext}})+\log k,  \label{PRRE Haw}
    \end{split}
\end{equation}
with the R\'enyi entropy $S_n=\frac{1}{1-n} \log [\text{Tr}(\rho^n)]$. Therefore, we obtain the expression \eqref{RE Haw}.

\par Similarly, for the PRRE $S_n(\rho_{\text{ext}} || \sigma_{\text{WH}})$, we can follow the similar replica trick to peform the calculation, see FIG. \ref{replicageometry2}. Note that in this case, there is no combination of connected geometries as shown in FIG. \ref{disconnect}. Consequently, the factor $k^{-1}$ is not required. Namely, $\text{Tr}\big(\rho_{\text{ext}}^n \sigma_{\text{WH}}^m\big)= \text{Tr}\big(\rho_{\text{ext}}^n \big) \cdot \text{Tr}\big(\sigma_{\text{WH}}^m\big) $, then we have
\begin{equation}
    \begin{split}
        S_n(\rho_{\text{ext}} || \sigma_{\text{WH}}) &= \frac{1}{1-n}\log \bigg[\text{Tr}\big(\rho_{\text{ext}}^n\big) \cdot \text{Tr}\big(\sigma_{\text{WH}}^m\big) \bigg] \\
        &=-S_n(\rho_{\text{ext}})+ \frac{n}{n-1}S_{\text{BH}}^{(1-n)},  \label{PRRE WH}
    \end{split}
\end{equation}
where we use the relation $S_{\text{BH}}^{(1-n)}=\frac{1}{n} \log \bigg(\frac{Z_{1-n}}{Z_1^{1-n}} \bigg)$ and \eqref{tracesaddles}. 
\begin{figure}[htb]
\centering
\includegraphics[scale=0.3]{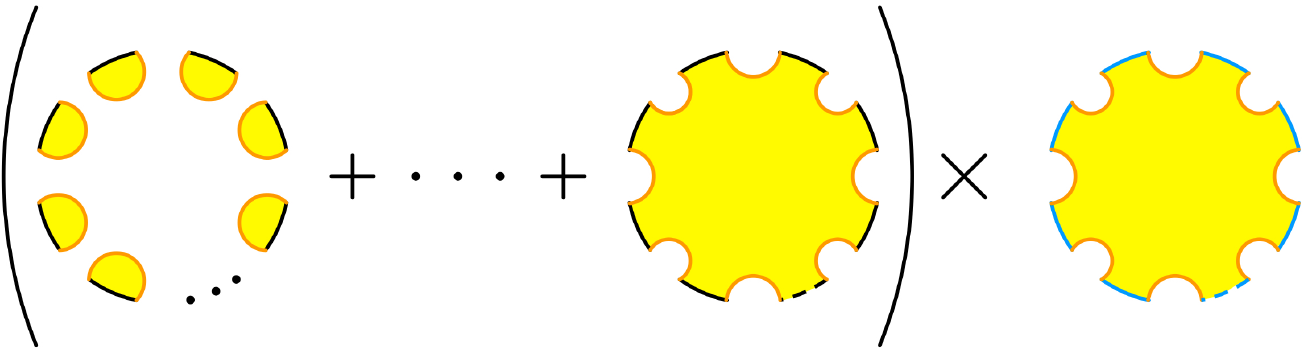}
\caption{Graphical representation of the reduced form of the trace $\text{Tr}\big(\rho_{\text{ext}}^n \sigma_{\text{WH}}^m\big)$.}
\label{replicageometry2}
\end{figure}

\par Now the final calculation is to take the limit of the above equation when $n \to 1$. The calculation hinges on two key steps: the exponential expansion of the matrix and the normalization of the density matrix. We start from the expression of the PRRE \eqref{PRRE WH}
\begin{equation}
    S_n(\rho_{\text{ext}}|| \sigma_{\text{WH}})= -S_n(\rho_{\text{ext}})+\frac{1}{n-1} \log \bigg[\text{Tr}(\sigma_{\text{WH}}^{1-n}) \bigg].  \label{PRRE WH2}
\end{equation}
The first term in the $n\to1$ limit is the von Neumann entropy: $S_n |_{n\to1}(\rho_{\text{ext}})=S_{\text{vN}}(\rho_{\text{ext}})$. So we need to expand the second term. Note that $\sigma_{\text{WH}}$ is the normalized density matrix
\begin{equation}
    \sigma_{\text{WH}}^1=\sigma_{\text{WH}}, \qquad \text{Tr} (\sigma_{\text{WH}})=1.  \label{normalized condition}
\end{equation}
In this time, we must use the weighted logarithmic formula. Setting $x=1-n$, then we have
\begin{equation}
    \begin{split}
        \text{Tr}(\sigma_{\text{WH}}^x) &= \text{Tr}\bigg( \sigma_{\text{WH}} \cdot e^{-x \log (\sigma_{\text{WH}})}\bigg )  \\
        &=\text{Tr}\bigg[\sigma_{\text{WH}} \bigg(I -x\log(\sigma_{\text{WH}})+\frac{1}{2}x^2\big(\log(\sigma_{\text{WH}}) \big)^2 +{\cal O}(x^2)\bigg)  \bigg] \\
        &=\text{Tr}(\sigma_{\text{WH}}) -x \text{Tr}\big(\sigma_{\text{WH}} \log(\sigma_{\text{WH}})\big)+{\cal O}(x^2)\\
        &=1-(1-n)\text{Tr}\big( \sigma_{\text{WH}} \log(\sigma_{\text{WH}})\big ) +{\cal O}(x^2). \label{expansion}
    \end{split}
\end{equation}
Thus, the second term in the $n\to1$ limit is reduced to
\begin{equation}
\begin{split}
    \frac{1}{n-1} \log \Big[ \text{Tr}\big(\sigma_{\text{WH}}^{1-n}\big) \Big] &\simeq \frac{1}{n-1} \log \Big[ 1+(n-1)\text{Tr} \big(\sigma_{\text{WH}} \log(\sigma_{\text{WH}}) \big) \Big]\\
    &= \text{Tr}\big( \sigma_{\text{WH}} \log (\sigma_{\text{WH}})\big) \\
    &= S_{\text{vN}}(\sigma_{\text{WH}}) \simeq S_{\text{BH}}.
\end{split}
\end{equation}
In the second line, we use the relation $\log(1+\varepsilon) \simeq \varepsilon$. The last equation is because $\sigma_{\text{WH}}$ represents the von Neumann entropy at late times, which is approximately equivalent to the BH entropy. Therefore, we can obtian the expression \eqref{RE WH}.

\par In conclusion, by applying the replica trick, we successfully derive the explicit expression for the RE \eqref{RE Haw} and \eqref{RE WH}.

\subsection{C. Explicit calculation for micro-canonical ensemble} 
\par In this appendix, we display a comprehensive calculation in the micro-canonical ensemble. The R\'enyi entropy of radiation in the PSSY model has the following form \cite{replica3}
\begin{equation}
\begin{split}
S^{\text{ME}}_n=S_0+\log t + 
\frac{1}{1-n} \log \Bigg[  \int d\lambda \cdot \lambda^n \cdot D(\lambda)\Bigg] &= \left \{
              \begin{array}{lr}  S_0 + \log t +\frac{1}{1-n} \log\Big[{_2}F_1 (1-n,-n,2,t)\Big], \quad  0\leq t\leq1. &\\
              S_0  +\frac{1}{1-n} \log\Big[{_2}F_1 (1-n,-n,2,\frac{1}t{})\Big], \qquad  \quad  \qquad  \ t\geq1. &
              \end{array} \right.  \label{renyi entropy}
\end{split}
\end{equation}
Here $\lambda$ and $D(\lambda)$ denotes the eigenvalue for $\rho_R$ and the density of state, respectively. Then for the $n\to1$ limit, we expand the hypergeometric function ${_2} F_1$  as follows
\begin{equation}
    \begin{split}
        t^{n-1} \cdot {_2} F_1 \Big(1-n,-n,2,\frac{1}{t} \Big)  &= {_2} F_1(1-n,-n,2,t)  \\
        &=1+(n-1)\frac{t}{2} - \frac{(n-1)^2}{2} \Big[1+\frac{3}{2}t + \frac{1-t^2}{t} \log (1-t) - 2\text{Li}_2(t) \Big] +{\cal O}\big[ (n-1)^2\big], \label{hypergeometric}
    \end{split}
\end{equation}
where $\text{Li}_2(t)$ refer to the polylogarithm function. Subsequently, the von Neumann entropy as the $n\to1$ case is written as 
\begin{equation}
\begin{split}
S^{\text{ME}}_\text{vN}=S_n^{\text{ME}}\big|_{n\to1}&= \left \{
              \begin{array}{lr}  S_0 + \log t -\frac{t}{2}, \quad  0\leq t\leq1. &\\
              S_0  - \frac{1}{2t}, \qquad  \quad  \qquad  \ t\geq1. &
              \end{array} \right.  \label{von ME}
\end{split}
\end{equation}
Substituting this expression into the expressions for RE \eqref{RE Haw} and \eqref{RE WH}, we can reproduce the expressions \eqref{RE Haw ME} and \eqref{RE WH ME}.

\subsection{D. Explicit calculation for canonical ensemble}
\par In this appendix, we provide a detailed derivation in the canonical ensemble and verified the numerical results in FIG. \ref{RE figures}  through some approximations. For the canonical ensemble with fixed $\beta$, we focus on the small $G_N$ region, namely, the semi-classical limit. Since $\beta=\beta G_N$ in the nature unit. The small $G_N$ region equivalent to the high-temperature limit with small $\beta$. In the beginning, in order to calculate the trace $\text{Tr}(\rho_{\text{ext}}^n)$, we can use the resolve matrix $R(\lambda)$. It can be solved by Schwinger-Dyson equation in the planar limit (large $k$)
\begin{equation}
    \frac{k}{R(\lambda)} + \int_0^{\infty} \frac{w(s) R(\lambda)}{k-w(s)R(\lambda)} = \lambda,  \label{SD eq}
\end{equation}
where the parameters are defined by \cite{replica1}
\begin{equation}
    \tilde{\rho}(s) = e^{S_0} \frac{s}{2\pi} \sinh(2\pi s), \quad Z_n= \int_0^{\infty} ds \cdot \tilde{\rho}(s) y^n(s), \quad y(s)= e^{-\frac{\beta}{2} s^2} \cdot2^{1-2\mu} \cdot \bigg| \Gamma\big(\mu-\frac{1}{2} +is  \big)\bigg|^2, \quad w(s) =\frac{y(s)}{Z_1}. \label{4parameters}
\end{equation}
For small $\beta$, the density of state $D(\lambda)$can be divided into two different stages in the evolution. Before the Page time, $D(\lambda)$ almost concentrates at $\lambda=\frac{1}{k}$. After this moment, however, $D(\lambda)$ gradually becomes a thermal spectrum, which reproduces a sharp cutoff at $s=s_k$ that is given by \cite{replica1}
\begin{equation}
    k= \int_0^{s_k} \tilde{\rho}(s) ds = e^{S_0} \frac{2\pi s_k \cosh(2\pi s_k)-\sinh(2\pi s_k)}{8\pi^4}. \label{cutoff}
\end{equation}
Then we calculate the minimal eigenvalue $\lambda_0$ of $D(\lambda)$ by $\frac{d\lambda}{dR}=0$: $\lambda_0\simeq \int_0^{s_k} ds\cdot \tilde{\rho}(s) w(s) = \frac{1}{k} \bigg(1- \int_0^{s_k} ds \cdot \tilde{\rho}(s) w(s) \bigg)$. The equation \eqref{SD eq} reduces to 
\begin{equation}
    (\lambda-\lambda_0) R \simeq k + \int_0^{s_k} ds \cdot \tilde{\rho}(s) \frac{w(s)R}{k-w(s)R}.  \label{SD eq2}
\end{equation}
Under the large $k$ limit $k \gg \int_0^{s_k} \tilde{\rho}(s) \frac{w(s)}{\lambda-\lambda_0-w(s)}$, we finally determine the density of state 
\begin{equation}
    D(\lambda) = \int_0^{s_k} \tilde{\rho}(s) \delta(\lambda-\lambda_0-w(s)), \label{dos}
\end{equation}
with two normalization conditions $\int_0^{\infty} d\lambda \cdot D (\lambda)=k$, and $\int_0^{\infty} d\lambda  \cdot \lambda \cdot D(\lambda)=1$. Accordingly, the von Neumann entropy is obtained by 
\begin{equation}
\begin{split}
        S_{\text{vN}}^{\text{CE}} &= - \int_0^{\infty}d\lambda \cdot D(\lambda) \ \lambda \log \lambda \\
        &=-\int_0^{s_k} ds \cdot \tilde{\rho}(s) \big(\lambda_0 +w(s) \big ) \log \big(\lambda_0+w(s) \big).  \label{vn CE}
\end{split}
\end{equation}
Through the REs \eqref{RE Haw} and \eqref{RE WH}, the expressions \eqref{RE Haw CE} and \eqref{RE WH CE} can be derived.

\par Now we use some suitable approximations to present the analytical  form for REs \eqref{RE Haw CE} and \eqref{RE WH CE} and employ it to validate the accuracy of FIG. \ref{RE figures}. At early stage, $k$ is sufficiently small, which leads to $D(\lambda)$ exists a sharp peaked around $\frac{1}{k}$ \cite{replica1}
\begin{equation}
    D^{\text{early}}(\lambda) \simeq \int_0^{s_k} ds \cdot \tilde{\rho}(s) \delta\big( \lambda-\frac{1}{k} \big ). \label{early D}
\end{equation}
Correspondingly, the von Neumann entropy \eqref{von ME} approximates to 
\begin{equation}
    \begin{split}
        S_{\text{vN}}^{\text{early}} &= -\int_0^{\infty} d\lambda \cdot D^{\text{early}}(\lambda) \cdot \lambda\log\lambda \\
        &\simeq -\int_0^{\infty} d\lambda \cdot D^{\text{early}}(\lambda) \bigg( \frac{1}{k}+ \bigg(\lambda-\frac{1}{k}\bigg)\bigg) \bigg[\log k - k \bigg(\lambda-\frac{1}{k}\bigg)  +\frac{k^2}{2} \bigg(\lambda-\frac{1}{k}\bigg)^2 +{\cal O}\bigg( \bigg(\lambda-\frac{1}{k}\bigg) \bigg)^3\bigg]. \label{early vn1}
    \end{split}
\end{equation}
Using the normalized conditions
\begin{equation}
    \int d\lambda \cdot D(\lambda)=k, \qquad \int d\lambda \cdot D(\lambda) \cdot \bigg( \lambda-\frac{1}{k} \bigg)=0,  \qquad \int d\lambda \cdot D(\lambda) \cdot \bigg( \lambda-\frac{1}{k} \bigg)^2=\frac{Z_2}{Z_1^2},
\end{equation}
where the partition function $Z_1\simeq1958.81$ and $Z_2\simeq1.32$ are defined by \eqref{4parameters}. Then the entropy is given by 
\begin{equation}
     S_{\text{vN}}^{\text{early}} \simeq \log k - \frac{1}{2}k \frac{Z_2}{Z_1^2} \sim \log k - k \cdot {\cal O}(10^{-7}).  \label{early vn2}
\end{equation}
So we therefore obtain the RE \eqref{RE Haw} at early times behaves as
\begin{equation}
\begin{split}
       S^{\text{early}} (\rho_{\text{ext}}|| \sigma_{\text{Haw}})  &=-S_{\text{vN}}^{\text{early}}(\rho_{\text{ext}}) + \log  k \\
       &\simeq k \cdot {\cal O}(10^{-7}) \gtrsim 0  \label{RE1 early}
\end{split}
\end{equation}
This result is consistent with FIG. \ref{RE figures}. The RE \eqref{RE WH} involves the BH entropy, we will discuss it in the following content. Nest we concentrate on the later stage, where $k$ becomes very large. We make the following approximations 
\begin{equation}
    \lambda_0 \sim \frac{1}{k} \simeq 0,  \qquad D^{\text{late}}(\lambda) \simeq \int_0^{s_k} \tilde{\rho}(s) \delta(\lambda-w(s)).
\end{equation}
For this case, the von Neumann entropy approximates to the BH entropy
\begin{equation}
\begin{split}
        S_{\text{vN}}^{\text{late}} &\simeq S_{\text{BH}} =- \int_0^{\infty} d\lambda \cdot D^{\text{late}}(\lambda) \cdot  \lambda \log\lambda \\
        &=- \frac{1}{Z_1} \int_0^{\infty} ds \cdot \tilde{\rho}(s) y(s) \log\big(y(s)\big)+\log Z_1.  \label{vn late}
\end{split}
\end{equation}
For $\mu \gg \frac{1}{\beta}$, $y(s) \sim y(0) e^{- \frac{\beta s^2}{2}}$ and $\big|\Gamma(\mu - \frac{1}{2}+is) \big|^2 /\big|\Gamma(\mu - \frac{1}{2}) \big|^2 = \prod_{m=0}^{\infty}(1+s^2)(m+\mu -\frac{1}{2})^2 \sim 1$. Then the above equation can be reduced to 
\begin{equation}
    S_{\text{BH}} \simeq  S_0+ \frac{4\pi^2}{\beta} +\frac{3}{2} \log \frac{2\pi}{\beta} + \frac{3}{2}- 2\log(2\pi) \simeq 20.285. \label{vn late2}
\end{equation}
Therefore, we can obtain the REs \eqref{RE Haw} and \eqref{RE WH} at early and late times as follows
\begin{equation}
    S^{\text{late}}(\rho_{\text{ext}}||\sigma_{\text{Haw}}) = - S_{\text{vN}}^{\text{late}}(\rho_{\text{ext}}) + \log k = \log k - \text{constant}  \sim \log k .  \label{RE1 late}
\end{equation}
\begin{equation}
      S^{\text{early}}(\rho_{\text{ext}}||\sigma_{\text{WH}}) =- S_{\text{vN}}^{\text{early}}(\rho_{\text{ext}}) +S_{\text{BH}} \simeq S_{\text{BH}}-\log k \gg 0.  \label{RE2 early}
\end{equation}
\begin{equation}
    S^{\text{late}}(\rho_{\text{ext}}||\sigma_{\text{WH}}) =- S_{\text{vN}}^{\text{late}}(\rho_{\text{ext}}) +S_{\text{BH}}\sim 0. \label{RE2 late}
\end{equation}
In conclusion, we can prove the correctness of the behavior of REs in FIG. \ref{RE figures} (with $\log k$ as the variable) through equations \eqref{RE1 early}, \eqref{RE1 late}, \eqref{RE2 early}, and \eqref{RE2 late}.

\end{widetext}

\begin{thebibliography}{99}
		
\bibitem{IL} S. W. Hawking, ``Breakdown of Predictability in Gravitational Collapse,''  Phys. Rev. D \textbf{14}, (1976) 2460-2473. 

\bibitem{HR} S. W. Hawking, ``Particle Creation by Black Holes,'' Commun. Math. Phys. \textbf{43}, 199 (1975). Erratum;[Commun. Math. Phys. \textbf{46}, 206 (1976)].

\bibitem{holography} J. Maldacena, ``The large N limit of superconformal field theories and supergravity,'' Int. J. Theor. Phys. \textbf{38}, 1113–1133 (1999). arXiv:hep-th/9711200.

\bibitem{islandrule1} A. Almheiri, N. Engelhardt, D. Marolf and H. Maxfield, ``The entropy of bulk quantum fields and the entanglement wedge of an evaporating black hole,'' JHEP \textbf{12}, 063 (2019) arXiv:1905.08762.

\bibitem{islandrule2} A. Almheiri, R. Mahajan, J. Maldacena and Y. Zhao, ``The Page curve of Hawking ridiation from semiclassical geometry,'' JHEP \textbf{03}, 149 (2020) arXiv:1908.10996.

\bibitem{replicawormholes} G. Penington, ``Entanglement Wedge Reconstruction and the Information Paradox,'' JHEP \textbf{09} 002 (2020) arXiv:1905.08255.

\bibitem{review} A. Almheiri, T. Hartman, J. Maldacena, E. Shaghoulian and A. Tajdini, ``The entropy of Hawking radiation,'' Rev. Mod. Phys. \textbf{93}, 35002 (2021) arXiv:2006.06872.

\bibitem{pc1} D. N. Page, ``Information in black hole radiation,'' Phys. Rev. Lett. \textbf{71}, 3743-3746 (1993) arXiv:hep-th/9306083.

\bibitem{pc2} D. N. Page, ``Time Dependence of Hawking Radiation Entropy,'' JCAP \textbf{1309}, 028 (2013) arXiv:1301.4995.

\bibitem{zeno} X. Ge, ``Zeno's paradox and black hole information loss problem,'' Acta Phys.Sin. \textbf{74}, (2025) 081101 arXiv:2511.10106.

\bibitem{replica1} G. Penington, S. Shenker, D. Stanford and Z. Yang, ``Replica wormholes and the black hole interior,'' JHEP \textbf{03}, 205 (2022) arXiv:1911.11977.

\bibitem{replica2} A. Almheiri, T. Hartman, J. Maldacena, E. Shaghoulian and A. Tajdini, ``Replica Wormholes and the Entropy of Hawking Radiation,'' JHEP \textbf{05}, 013 (2020) arXiv:1911.12333.

\bibitem{cptp} G. Lindblad, Completely positive maps and entropy
inequalities, Commun. Math. Phys. \textbf{40}, 147 (1975).

\bibitem{crooks} G. E. Crooks, Phys. Rev. E \textbf{60}, 2721 (1999).

\bibitem{jtgravity1} C. Teitelboim, ``Gravitation and Hamiltonian Structure in Two Space-Time Dimensions,'' Phys. Lett. B \textbf{126}, (1983) 41.

\bibitem{jtgravity2} R. Jackiw, ``Lower Dimensional Gravity'', Nucl. Phys. B \textbf{252}, (1985) 343.


\bibitem{replica3} M. Yu, S. Lin, and X. Ge, ``Replica wormholes, modular entropy, and capacity of entanglement in JT gravity,'' Sci.China Phys.Mech.Astron. \textbf{69} (2026) 2, 220412 arXiv:2501.11474.


\bibitem{syk1} J. Maldacena and D. Stanford, ``Remarks on the Sachdev-Ye-Kitaev model,'' Phys. Rev. D \textbf{94} (2016) 106002 arXiv: 1604.07818.

\bibitem{syk2}  S. Sachdev and J. W. Ye, ``Gapless spin-fluid ground state in a random quantum Heisenberg magnet,'' Phys. Rev. Lett. \textbf{70}, (1993) 3339 arXiv:cond-mat/9212030.

\bibitem{syk3} X. Ge and C. Zhang, ``On the Replica Problem in Supersymmetric SYK Models,'' arXiv:2508.09611.



\end{thebibliography}
\end{document}